\documentclass[aps,prd,twocolumn,superscriptaddress]{revtex4}

\usepackage{graphicx}
\begin{document}

\title{Trapping of strangelets in the geomagnetic field}

\author{L. Paulucci}
\email[]{paulucci@fma.if.usp.br} \affiliation{Instituto de F\'\i
sica - Universidade de S\~ao Paulo\\
Rua do Mat\~ao, Travessa R, 187, 05508-090, Cidade Universit\'aria\\
S\~ao Paulo SP, Brazil}
\author{ J. E. Horvath }
\affiliation{Instituto de Astronomia, Geof\'\i sica e Ci\^encias
Atmosf\'ericas - Universidade de S\~ao Paulo\\
Rua do Mat\~ao, 1226, 05508-900, Cidade Universit\'aria\\
S\~ao Paulo SP, Brazil}
\author{ G. A. Medina-Tanco}
\affiliation{Instituto de Ciencias Nucleares, Universidad Nacional
Aut\'onoma de M\'exico\\
A.P. 70-543, C.U. M\'exico D.F., M\'exico}

\date{\today}

\begin{abstract}
Strangelets coming from the interstellar medium (ISM) are an
interesting target to experiments searching for evidence of this
hypothetic state of hadronic matter. We entertain the possibility
of a {\it trapped} strangelet population, quite analogous to
ordinary nuclei and electron belts. For a population of
strangelets to be trapped by the geomagnetic field, these incoming
particles would have to fulfill certain conditions, namely having
magnetic rigidities above the geomagnetic cutoff and below a
certain threshold for adiabatic motion to hold. We show in this
work that, for fully ionized strangelets, there is a narrow window
for stable trapping. An estimate of the stationary population is
presented and the dominant loss mechanisms discussed. It is shown
that the population would be substantially enhanced with respect
to the ISM flux (up to two orders of magnitude) due to
quasi-stable trapping.
\end{abstract}


\maketitle

\section{Introduction}
In a celebrated paper Witten \cite{Wit} elaborated on the
possibility \cite{Bodmer, Chin, Terazawa} that systems composed of
an deconfined Fermi gas of up, down and strange quarks could have
a lower energy per baryon than iron, thus being absolutely stable.
This hypothetical state (strange quark matter) could be created by
weak interactions introducing the massive $s$ quark, if the energy
``cost" of the mass is compensated by the availability of a new
Fermi sea associated to this extra flavor, thus lowering the Fermi
energy of the $u$ and $d$ quark seas.

Previous works have shown \cite{Farhi84} that this stability may
be realized for a wide range of parameters of strange quark matter
(SQM) in bulk on the basis of the MIT bag model. Calculations also
indicate that SQM can be absolutely stable within other models,
e.g. shell model \cite{Jaf, Madsen98}, or not stable at all
\cite{NJL}. More recently, studies have indicated that a paired
version of SQM, the CFL (color-flavor locked) state seems to be
even more favorable energetically than the unpaired SQM, widening
the stability window \cite{CFL2, CFL3, CFL4, CFL1}.

For the description of \textit{finite} size lumps of strange
matter, (termed \textit{strangelets}) a few terms have to be added
to the bulk one in the free energy. A surface term suffices for $A
\gg 10^{7}$, while other corrections are relevant for the lower
masses (see \cite{Mads} for a recent review). Large lumps will
have essentially the same number of quarks of bulk matter, with a
small depletion of the massive strange quark resulting in a net
positive charge. This is a feature also expected for small chunks
\cite{Jaf, Mads}, which thus resemble heavy nuclei.

In spite of theoretical controversies, it is generally agreed that
the ultimate SQM proof must be provided by experiments. The
experimental searches of strangelets started some 20 years ago and
have been reviewed recently in \cite{Klin, Finch}. In addition to
direct production of strangelets in heavy ion collisions
\cite{Ions1, Ions2, Ions3, Ions4, Ions5, Ions6, Ions7, Ions8, Arsen},
cosmic rays may contain primaries in this state of matter, which could
eventually be detected directly or as a deposition in terrestrial
matter \cite{derujula84, bruegger89, isaac98,Lu04}.

Several cosmic ray events have been tentatively identified in the
past as primary strangelets (initially the Centauro events and the
HECRO-81 experiment \cite{Data1,Data2,Data3}) for they present
features such as their high penetration in the atmosphere, low
charge-to-mass ratio and exotic secondaries \cite{Pol}. More
recently, at least one event recorded from the AMS-01 experiment
\cite{Boiko}, a mass spectrometer aboard the shuttle Discovery
during a 10-day flight in 1998, is being considered as possible
detection of a strangelets. While it is tempting to identify the
primary as a strangelet, the inevitable shell effects complicate
the analysis and preclude any firm conclusion as yet
\cite{PaulucciHorvath}. It is not clear until today to what extent
the anomalous events can be originated by ordinary primaries or
rather forcefully require a truly exotic origin.

Considering the question of existence of strangelets among cosmic
ray primaries, a few injection (production) scenarios have been
considered. Witten originally suggested the merging of compact
stars as a likely site \cite{Wit}. In principle, injection spectra
and the total mass in the galaxy may be calculated knowing the
rate of the events and the total ejected mass in each of them.
These estimates are subject to some caveats, for example, while
the number of merging systems has been revised upwards \cite{DP},
numerical work has shown that a substantial ejection of matter is
not guaranteed \cite{Wod} in a strange star-black hole system, and
the situation is unclear in the case of a fully relativistic SS-SS
system, which has only been partially addressed
\cite{Coalescencia} because the calculations had other goals. On
the other hand, strange matter formation in type II supernova
\cite{Mac} has been preliminarily explored and in these events a
small fraction of strange matter may be ejected. A numerical
analysis has shown that the possible quark matter component of
cosmic rays primaries is compatible \cite{Gustavo} with models in
which strangelets are ejected in either scenario.

While an uncertain flux from this ``contamination" of the ISM is
expected \cite{Madsen05}, we would like to discuss in this paper
another likely site to search for strangelets of cosmic origin.
Much in the same way heavy nuclei are present in the earth's
magnetosphere bouncing between magnetic mirror points, strangelets
could also become trapped in specific regions of the magnetosphere
and their number density increased respect to the ISM flux,
provided some conditions for their capture by planetary magnetic
fields are met. This phenomenon is analogous to the Van Allen
belts, and has been first suggested in a former study \cite{Rosu}.
A handful of experiments have probed the magnetosphere by
measuring the fluxes of the so-called ``anomalous" cosmic ray
nuclei, and may already place interesting limits to strangelets as
well. Overall the existence and nature of exotic primaries is an
important issue. In addition to former and ongoing searches, there
will be a mass spectrometer placed at the International Space
Station, the AMS-02 experiment \cite{AMS, AMS2}, with one of its
goals to help the identification of this exotic component, of
crucial importance in testing the validity of the
Bodmer-Witten-Terazawa conjecture. We substantiate below the
strangelet belt idea, discuss the main features of this
population, and advocate for a search of this exotic component at
definite sites within existing uncertainties based on these
calculations.

\section{States of ionization and electronic recombination of strangelets in the ISM}

As is well-known, unpaired (also referred as ``normal" in this
work) SQM in bulk contains light $u$, $d$ and massive $s$ quarks
in $\beta$-equilibrium. Because of the depletion of the more
massive $s$ quark, a small fraction of electrons is also present
to maintain charge neutrality. On the other hand, SQM in a paired
CFL state is automatically neutral, since the equal number of
flavors is enforced by symmetry \cite{Rajagopal}. Actually, a
small positive charge is present because of the smaller abundance
of $s$ quarks near the surface in CFL strangelets \cite{Mads}.
Therefore it is natural that CFL strangelets will be surrounded by
an electronic cloud in order to neutralize its total charge,
forming an exotic atom. The same happens for normal strange matter
if the strangelet radius is smaller than the electron Compton
wavelength, a condition satisfied whenever $A \ll 10^{7}$.

In the following and throughout the whole analysis presented here,
the strangelet rest mass will be assumed to be $\epsilon_{0} \, \,
A \, \sim \, (930 \, \times \, A)$ MeV, with $\epsilon_{0}$ the
asymptotic value of the energy per baryon of strange quark matter.
We will not consider the fact that the energy per baryon number
decreases with $A$ in sophisticated model calculations, given that
the uncertainties found in other parameter choices are expected to
be much larger than the error associated with this approximation.
Also the strange quark mass is considered to be $m_{s} \, = \, 150
\, MeV$ and the coupling gap of CFL strange quark matter, $\Delta
\, = \, 100 \, MeV$ in this exploratory study. With these
assumptions, the net positive charge of strangelets is given
approximately by $Z=0.1 \, A$ (low baryon number regime) in the
MIT bag model approach for normal strange matter and $Z=0.3 \,
A^{2/3}$ for the CFL model.

Strangelets from whatever astrophysical injection event would
travel through the interstellar medium and become ionized by
collisions. A simple analysis to evaluate the degree of ionization
of semi-relativistic strangelets surrounded by electronic clouds
due to these interactions was performed in a Bohr atom
approximation. Strangelets are partly neutralized by electrons
from the excitation of the vacuum if $Z \, \gg \, 100$
\cite{Madsen03}, but for all cases of interest in this work the
baryon number range is such that we do not have to deal with this
effect.

We considered a two-body collision (incident electron - electron in
the strangelet cloud) instead of a multibody problem, which would be
much more difficult to handle. The stripping interactions are mainly
due to electrons with a Maxwellian speed distribution at a
temperature of $\sim \, 100 \, K$, an average condition of electrons
in the ISM.

\begin{figure}
\includegraphics{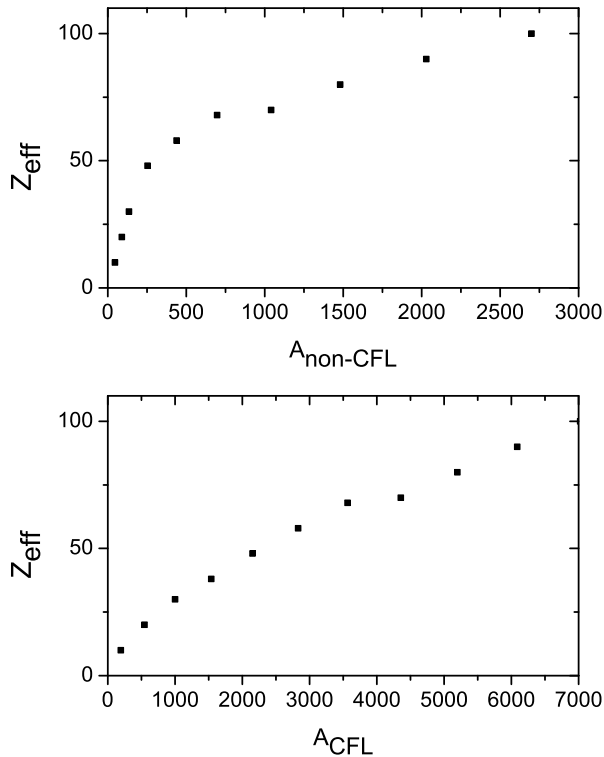}
\caption{Strangelet effective charge ($1 \, GeV/A$) versus the
baryon number $A$ for normal (upper panel) and CFL (lower panel) strange matter
after interaction with electrons in the interstellar medium.}\label{Zeff}
\end{figure}

The results are shown in figure \ref{Zeff} for strangelets with
total energy of $1 \, GeV/A$. Considering the average density in the
interstellar medium to be $1 \, particle/cm^{3}$, the mean free path
for an electronic collision which may or may not result in
ionization is of the order of $10^{15} \, cm$, which is very short
on astronomical standards.

The ionization degree became stable within a travelled distance of
a few $pc$ for $1 \, GeV/A$ strangelets. For ultra-relativistic
strangelets (i.e., of the type of candidates that would produce a
Centauro event  \cite{Lattes} $E/A \, \sim \, TeV$) the
calculations indicate always full ionization. Furthermore,
according to the model proposed by Werner and Salpeter
\cite{Werner} for the radiation flux in the ISM, the influence of
the radiation field on ionization of strangelets will be
negligible unless the strangelet trajectory crosses a region
containing very energetic photons (i.e. the surroundings of a
Wolf-Rayet, O and B stars and/or regions of stellar formation).

We acknowledge that a Bohr atom treatment is a crude approach for
the electron distribution around the strangelet, since it does not
include quantum corrections as important as the spin-orbit
coupling and non-local effects, nor relativistic corrections for
many electrons bodies ($Z \, \geq \, 40$). There is no general
expression for these corrections applicable in the case of atoms
with many electrons though; the existent models (e.g.,
Hartree-Fock calculations) are restricted to atoms with few
electrons, the same happening for experimental corrections. In
this way, the calculations presented here are rough estimates,
showing the general trend of the effects rather than providing
precise numerical values.

For low-energy particles the electronic capture can be as
important as the ionization process thus far discussed. An
approximate cross section for the capture of electrons of velocity
$v$ by a charged particle of atomic number $Z$ is given as
\cite{Massey} \cite{Bethe}

\begin{equation}\label{capture}
\sigma_c=Z^22^{2/3}\alpha^4\frac{h^2\nu^2}{m_e^2v^2c^2}\Big(\frac{m_ec^2}{h\nu}\Big)^{7/2}\times
6.65\times 10^{-25}\, cm^2,
\end{equation}
\\
where $h\nu\approx E_e$ for $E_e>>I$, $I$ and $E_e$ being the
electron energies while bound to the nucleus and free in the ISM,
respectively, and $m_ec^2$ is the electron rest mass. This form of
the cross-section for radiative recombination is obtained relating
the capture of a bare nucleus of charge $Ze$ with the capture into
the corresponding state of a hydrogen atom, which is proportional
to the energy of the gamma emitted in the process and also to the
cross-section for the absorption of a quantum of frequency $\nu$
by a $H^-$ ion resulting in emission of an electron of velocity
$v$, and inversely proportional to the momentum of the electron
absorbed. In the case of a partially screened nucleus, the
cross-section is still given approximately by equation
(\ref{capture}), though a special calculation must be performed to
obtain the cross-section for capture into an orbital with quantum
number $n_0$, usually given in tables for ordinary nuclei.

The ``atom'' or ``ion'' formed by capturing an electron may also
lose this electron in further interactions. For light materials
the cross section for electron loss can be approximately expressed
for $v > v_0$ \cite{Bohr48} as

\begin{equation}
\sigma_l=8\pi a_0^2Z^{-2}\Big(\frac{v_0}{v}\Big)^2,
\end{equation}
\\
where $a_0=\hbar^2/me^2=0.53\times 10^{-8}$ cm is the Bohr radius
and $v_0=e^2/\hbar$, whereas for intermediate $Z$ materials

\begin{equation}
\sigma_l=\pi a_0^2Z^{-1}\Big(\frac{v_0}{v}\Big),
\end{equation}
\\
because of the screening effect.

In summary, a comparison of eqs.(1), (2) and (3) shows that
electronic capture would only be important for high Z strangelets,
precisely where this simple picture can no longer be applied due
to vacuum excitation effects. That corresponds to a region
in baryon number which we believe to be of minimum relevance to the trapped population.

In summary, these results indicate that we can assume total
ionization as a good approximation to incoming ISM strangelets
that could form an ionization belt in the magnetosphere.

\section{Capture of strangelets in the geomagnetic field}

The motion of ionized strangelets in the earth magnetosphere can be
studied by applying the St\"ormer theory in a dipolar magnetic
field. The movement analysis can be made in terms of the geomagnetic latitude
and the \textit {L} parameter, where $L$ is the equatorial distance of a
field line to the axis of the dipole measured in units of the earth
radius.

The geomagnetic field is not a pure dipole field. Instead, most
magnetic models used for studying it include nearly 50 terms for
describing the potential field from which the magnetic field is
obtained in a sum of Legendre functions multiplied by oscillatory
coefficients in the azimuthal variable. Since the potential field
has a $r^{-(n+1)}$ dependence, the importance of high-order terms
decreases rapidly as one moves away from the earth surface. In
this way, the $n=1$ term, i. e., the dipole term, is the lowest
but dominant term, and most features of the trapped radiation
theory are analyzed based on a dipole field.

Charged particles with energy of order of $MeV$ in the inner part
of the magnetosphere ($L \ll 10$) rotate with a much higher
frequency than that of typical geomagnetic field variation (which
varies in time scales of, at most, few minutes). Under these
conditions, the magnetic moment is a conserved quantity (adiabatic
invariant). Therefore particles with high enough magnetic moments
become trapped in the dipolar field lines of the geomagnetic
field, with mirror points placed near the earth poles.

Particles with mirror points that allow penetration in the earth
atmosphere can be lost via collisions with atoms. All the
particles with mirror points placed inside the earth radius are
obviously lost, meaning that particles with
$|\alpha_{eq}|<\alpha_E$ or $|\pi-\alpha_{eq}|<\alpha_E$, where
$\alpha_{eq}$ is the equatorial pitch angle, are inside the earth
loss cone.

We will consider collisions mainly with the neutral nitrogen
molecule ($N_2$). The probability of interaction of trapped
particles which penetrate the atmosphere (suffering collisions
losses) can be taken as

\begin{equation}
P(s)=1-e^{-s/\lambda(s)}
\end{equation}
\\
at a certain point $s$, since each process is probabilistically
independent, being $\lambda$ the particle mean free path.
Generalizing the previous equation, it is necessary
to integrate over the particle path. Assuming that all the
strangelets which collide with particles in the atmosphere are
eventually removed from the trapped flux, we express the escape
probability as

\begin{equation}
P_{esc}=1-e^{-\int_{s}\sigma\Big[n(s')+s'\frac{dn}{ds'}\Big]ds'}
\end{equation}
\\
where $ds=LR_E\cos\lambda\sqrt{1+3\sin^2\lambda}d\lambda$ is the
arc along a field line, $\sigma$ is the particle cross-section
and $n(s)$ is the density of particles in
the atmosphere at a certain point $s$ of the strangelet's path.
Since strangelets are hadrons we may take their relevant
interaction cross-section to be geometrical ($\sigma \propto \,
A^{2/3}$).

\begin{figure}
\includegraphics[width=0.5\textwidth]{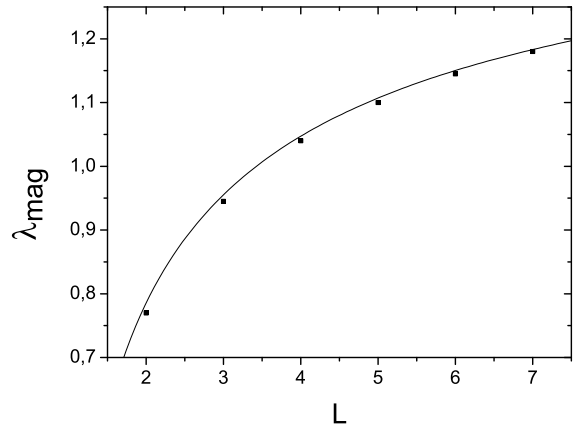}
\caption{Loss cone for strangelets in the geomagnetic field, i. e.,
particles with mirror point placed at geomagnetic latitudes above
the square dots (coming from the analysis of collisions with
atmospheric particles) are eventually removed from the trapped
population. The full line represent mirror points at the earth
surface in the dipole field approximation.}\label{stability}
\end{figure}

The calculated loss cone for strangelets, assuming an exponential
profile of the atmospheric density is shown in figure
\ref{stability} for different $L$ reflecting collisions with
atmospheric particles and the non-existence of a suitable mirror
point. It indicates, as expected, that the smaller the equatorial
pitch angle the easiest it is to remove a trapped particle.

In order for a particle to penetrate a certain region in the
magnetosphere its energy must be enough to overcome the local
geomagnetic cutoff rigidity. A solution was found by St\"ormer \cite{Stormer}
describing a special case of what he called the ``forbidden cone'', which
gives the geomagnetic cutoff rigidity. In this way, the condition a particle
must fulfill to have access to a given region of the magnetosphere
can be written as \cite{Cut}

\begin{equation}\label{cutoff}
R_{particle} \, > \,
\frac{59.6\,\cos^4\lambda}{L^{2}[1+(1-\cos\gamma\cos^3\lambda)^{1/2}]^2}\,\,\,\,GV
\end{equation}
\\
where $\lambda$ is the latitude and $\gamma$ the arrival direction of the
particle (east - west).

In the analysis of charged particles trapped in a magnetic field it is usually
considered that the motion of a given particle is a composition of three
different motions: the bouncing motion of a guiding center along the magnetic
field line; the rotational motion of the particle itself around that guiding
center; and the longitudinal drift of the guiding center. In this way, the
condition for a $triply-adiabatic$ motion is that the magnetic field intensity
must vary very slowly around a cyclotron orbit, imposing a \textit{maximum}
energy allowed for stable trapping. The condition that
must be imposed for the cyclotron radius at the equator
is given by

\begin{equation}\label{rho}
R_C \,\Big |_{equator} = \, \frac{p_{\perp}}{qB} \, \ll \,
\frac{B}{|\nabla_{\perp}B|}\Big |_{equator}
\end{equation}

\begin{figure*}
\includegraphics[width=0.8\textwidth]{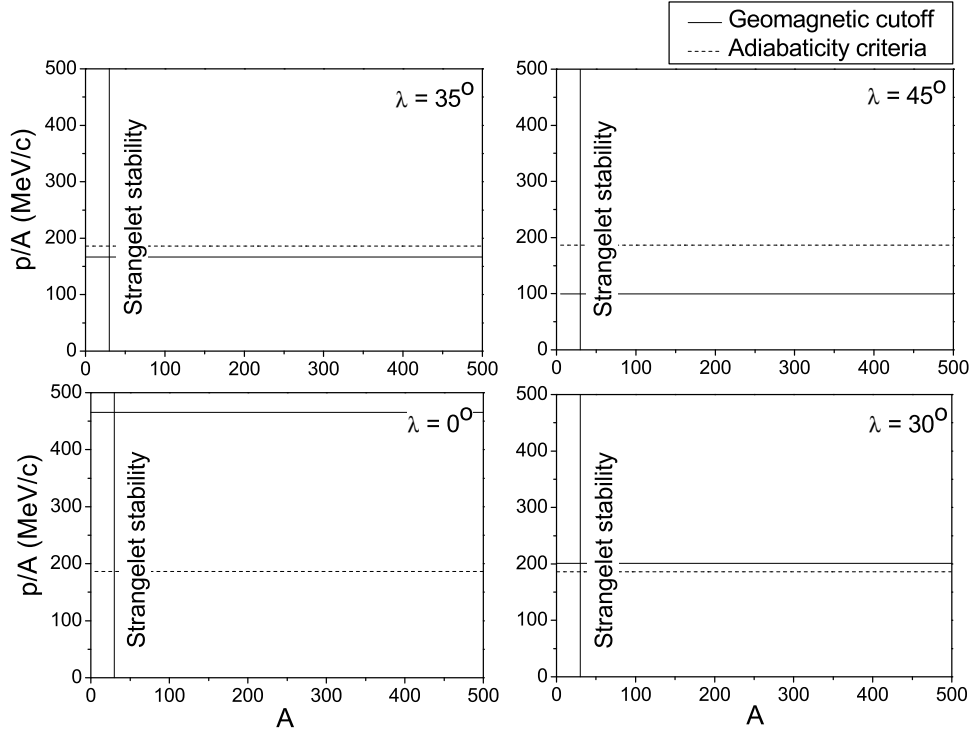}
\caption{Restriction curves (\ref{cutoff}) and (\ref{rho}) for $L=2$
in the baryon number vs. momentum plane for normal strangelets
incident from the east ($\gamma=\pi$) for different incident
directions (reminding that field lines at $L=2$ penetrate the earth
surface at $\lambda=45^{o}$ in the dipole model).}\label{L2NSQM}
\end{figure*}

\begin{figure*}
\includegraphics[width=0.8\textwidth]{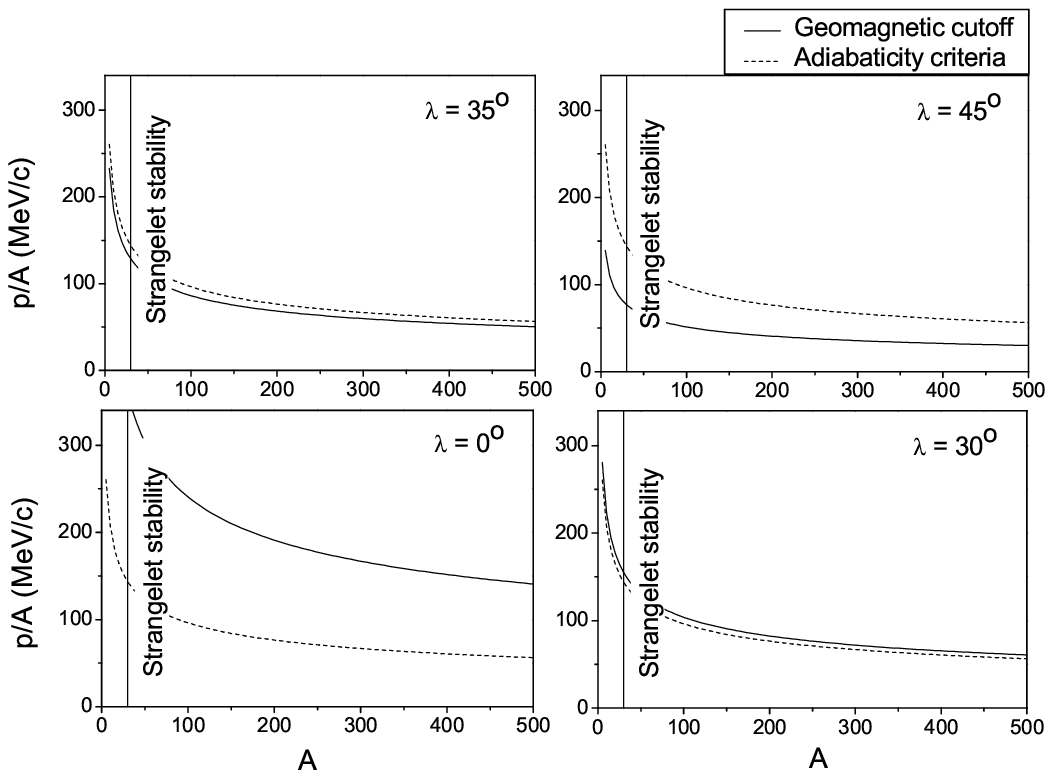}
\caption{The same as Figure \ref{L2NSQM} for CFL
strangelets.}\label{L2CFL}
\end{figure*}

Figures \ref{L2NSQM} and \ref{L2CFL} show those bounds for normal
and CFL strangelets, respectively, for $L \, = \, 2$ in addition
to the minimum baryon number which is required for strangelet
stability \cite{Madsen98}. The existence of a minimum baryon
number is expected in all models of SQM because the energy needed
for producing the system increases as the baryon number decreases,
till it reaches a value above which the strange matter is
unstable. The value adopted has been $A_{min} \, = \, 30$ (shown
with a vertical line) and may be trivially altered for any other
figure. Strangelets with very high baryon number, though allowed
for stable trapping, are not likely to be statistically
significant for detection in the magnetosphere due to a
substantial decrease of the interstellar flux expected as the
baryon number increases.

The upper bound (\ref{rho}) has been enforced in our calculations
to a 10\% confidence level according to observations of anomalous
cosmic rays $L$-shell distributions \cite{Tylka}, and we
considered $E_{\perp}\sim E$, which means we are actually
\textit{underestimating} the number of particles that could be
stably trapped in the geomagnetic field. Obviously, the
geomagnetic cutoff curve must be below the adiabaticity criteria
for stable trapping to occur. This is not the case for small
latitudes, but there is a narrow ``window" in latitude starting
slightly above $30$ degrees at $L=2$ for strangelets from the
ISM flux to fulfill the conditions of capture and accumulate in
regions labelled by the $L$ parameter. However, the number of
accumulated particles is still interesting, as shown in the next
section.

When this calculation is repeated for the case of CFL strangelets,
the region allowed for stable trapping for CFL strangelets has a
different shape than that for normal strangelets. This feature is
due to the strong dependence of the charge upon $A$ of normal
strangelets ($Z \, \propto \, A$) resulting in constant values if
one considers momentum per baryon number, whereas for the CFL
strangelets charge is almost independent of $A$, leading to a
$\sim A^{-0.9}$ dependence in the momentum per baryon number
variable.

\subsection{Trapped strangelet population}

Even though strangelets can be captured and trapped in the earth's
magnetic field, we must evaluate the possible maintenance of a
strangelet population to check whether there is an increase of the
flux. For this purpose, we must consider losses mechanisms.

In addition to the already analyzed losses by collisions with
neutral atmospheric particles, we have considered the {\it inward
drift} driven by asymmetric fluctuations of the geomagnetic field
as a dominant mechanism to diminish the strangelet population.

We will not consider in this work direct pitch angle diffusion.
Because of their large mass, strangelets are less likely to be
scattered appreciably in pitch angle by collisions. The net result
of multiple collisions with atmospheric particles would be a
reduction in the strangelets kinetic energy to thermal values and
minor changes in their pitch angle. Since we have already
considered that particles bouncing at a radial distance from the
surface of the earth below the atmosphere height scale (derived in
section II) would be eventually removed, we are in fact replacing
a diffusion equation in the $\cos(\alpha_{eq})$ variable for a
constant loss term (a sink function) directly related, though not
formally assigned, to pitch angle diffusion.

Radial diffusion must proceed by fluctuations in the third
invariant $\phi$, which is proportional to $L^{-1}$, due to
changes in the electric or magnetic fields that are more rapid
than the particle drift frequency. Because the gyration and bounce
periods are much shorter than the drift period, the first and
second adiabatic invariants are less likely to be affected by many of these
field perturbations.

Guided by the existing calculations and observations for anomalous
cosmic ray nuclei (hereafter ACR) trapping, we have considered third
invariant diffusion due asymmetric fluctuations in the geomagnetic
field, which is mainly driven by the solar wind pressure (sudden
compression and slow relaxation of the geomagnetic field).

The diffusion coefficient $D_{LL}$ is determined theoretically by 
taking two consecutive steps \cite{Walt}. First, one has to
evaluate the radial displacement suffered by a particle under the
influence of the field disturbance, which is an idealized model of the
real disturbances occuring in the geomagnetic field. 
The following procedure is taken in order to obtain
the diffusion coefficient as a function of the statistical features
of the disturbances alone. It consists of squaring this displacement and 
taking the average over several
disturbances randomly occurring in time and over all possible particle's
initial longitudes.

The diffusion coefficient due to magnetic field fluctuations for
equatorially trapped particles, with the assumption of efficient
phase mixing \cite{Walt} can be expressed as

\begin{equation}\label{DLL}
D_{LL}^M=\frac{\pi^2}{2}\Big(\frac{5}{7}\Big)^2\frac{R_E^2\,L^{10}}
{B_0^2}\nu_{drift}^2\,P_A(\nu_{drift})
\end{equation}
\\
where $P_A(\nu)$ is the power spectral density of the field
variation evaluated at the drift frequency. For off-equatorial
particles, the diffusion coefficient presents an exponential decay
with latitude.

Already in the case of nuclei, it is known that the complex
geometry and inhomogeneities in the geomagnetic field make
quantitative calculations ambiguous. The observed values of the
diffusion coefficient and their $L$ dependence will change with
global magnetic activity, and magnetic disturbances are known to
vary appreciably with time. We have assumed a $\nu^{-2}$
dependence of the power spectral density for simplicity
\cite{Walt}. The loss of more detailed information associated with
this approximation is that the diffusion coefficient becomes
independent of the energy of the particle entering the geomagnetic
field. In this case, the diffusion coefficient have a strong
dependence on the {\it McIlwain parameter} ($D_{LL}\propto
L^{10}$) \footnote{Radial diffusion caused by random variations in
the potential electric fields have a softer dependence on the
McIlwain parameter, being the resultant diffusion coefficient
$D_{LL}\propto L^{6}$ \cite{electric}.}. This indicates that its
influence is very important for particles trapped at higher
$L$-shells.

Typical values for changes in the trapped population
distribution ranges from a few hours at $L=6$ to hundreds of days
at $L=2$. Therefore if strangelets are captured by the geomagnetic
field their density must be higher for lower values of the $L$
parameter, which may result in a substantial increase of this
population compared to the ISM flux.

Some other losses mechanisms are of less importance in short time
scales, but have influence on long time scales, thus lowering the
residence time for trapped particles. This includes electrical
drift-resonant interactions between particles and fields,
especially in the pulsation frequency or VLF range
\cite{Baumjohann}. Those phenomena are highly affected by the
solar wind activity.

The diffusion equation has been employed to study the trapped
strangelet flux

\begin{equation}\label{diff}
\frac{\partial f(\mu,J,L)}{\partial t}=\frac{\partial}{\partial
L}\Big[\frac{D_{LL}}{L^2}\frac{\partial}{\partial L}(L^2f(\mu,
J,L))\Big]
\end{equation}
\\
where $f$ is the distribution function, $D_{LL}$ is given by
equation (\ref{DLL}) and $\mu$ and $J$ are the adiabatic
invariants magnetic moment and integral invariant, respectively.
The relation between the distribution function and the flux may
be given by $j(E,\alpha)=p^2L^2f(\mu, J, L)$. A
stationary population requires $\partial f/\partial t=0$, i.
e., the assumption that the source and loss terms are
instantaneously balanced is valid.

\begin{figure}
\includegraphics[width=0.5\textwidth]{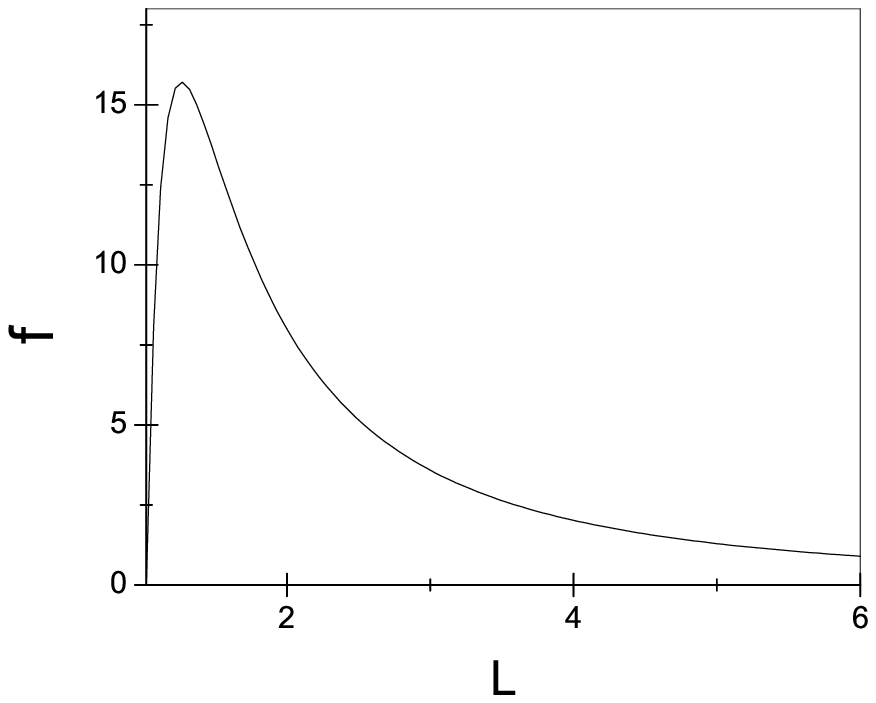}
\caption{The distribution function (in arbitrary units) for
strangelets trapped in the geomagnetic field as a function of $L$ is
obtained as the solution for the differential equation \ref{diff}
with the boundary conditions $f(L_{max})$ given by the incoming flux
(see text for details) and $f(L_{min})=0$ (corresponding to the
height scale of interaction with atmospheric particles). The
position of the peak (around $L=1.3$) does not change with the
change in the $A/Z$ relation (CFL and non-CFL strangelets) nor with
a change in the energy and A of the strangelets.}\label{density}
\end{figure}

We assume a steady strangelet injection from the interstellar
medium at $L=6$ (the position of the maximum distribution function
is very insensitive to the chosen L-shell parameter for this
boundary condition) and derive the distribution function shape
between this maximum and $L\approx 1.05$ where it is null
(atmosphere particle interaction height), shown in Figure
\ref{density}. We are not considering diffusion in pitch angle due
to interaction of particles with electromagnetic waves caused by
field variations, which alters the first adiabatic invariant.

The calculations were carried on with two values of the flux from
the ISM reaching the outer magnetosphere.

The first one, which will be called ``standard", is the one that
assumes the standard cosmic ray dependence on the strangelet flux,
$E^{-2.5}$. The total ISM strangelet flux that reaches the earth
as estimated by Madsen \cite{Madsen05} for a binary strange star
system coalescence scenario is given by

\begin{widetext}
\begin{equation}\label{flux}
F\approx 2 \times 10^5\,
m^{-2}\,yr^{-1}\,sr^{-1}\,A^{-0.467}\,Z^{-1.2}\,max[R_{SM},R_{GC}]^{-1,2}\Lambda
\end{equation}
\end{widetext}

where $R_{SM}$ and $R_{GC}$ are the solar modulation and geomagnetic
cutoff rigidities, respectively, and $\Lambda$ is an uncertain
parameter assumed to be of $O(1)$. In this way, the whole flux is fitted
with a $E^{-2.5}$ dependence with the constraints of minimum and maximum
energy respecting the values $R_{min}=5 MV A/Z$ and $R_{max}=10^6 GV$ \cite{Madsen05}.

The second calculation, which will be called ``improved",
considers a more detailed characterization of the differential
flux, where for the region of interest in this work (rigidities of
few GV), the strangelet flux actually {\it increases} with a slope
of $R^{1.8}$. This flux was obtained from a fit to reference
\cite{Madsen05}.

In either way, the flux entering the region of the magnetosphere
at $L_{max}$ has to fulfill the restrictions imposed for stable
trapping in the pitch angle and geomagnetic latitude of incidence

\begin{equation}
F_{in}=\int_{\lambda_{min}}^{\lambda_{max}} d\lambda\, P(\lambda)\,
\int_{\alpha_{loss\,cone}}^{\pi/2} d\alpha_{eq}\, P(\alpha_{eq}) \times
F_{}
\end{equation}
\\

The efficiency factors, $P(\lambda)$ and $P(\alpha_{eq})$ may be
easily identified: $P(\lambda)$ gives the fractional area of the spherical
section suitable for trapping discussed previously

\[
P(\lambda)=\frac{2L^2\,(-cos\lambda)\int_{0}^{2\pi}d\phi}{2L^2\,\int_{0}^{\pi/2}cos\theta\int_{0}^{2\pi}d\phi}
\]
\\
where the factor $2$ comes from the symmetry in $\theta$ for both
hemispheres (north, south). $P(\alpha_{eq})$ limits the number of
particles entering the specific region of the magnetosphere with
an appropriate pitch angle to avoid the loss cone as already
discussed. We have also assumed an isotropic flux of particles,
since there is no theoretical prediction pointing to any
anisotropy in the arrival direction of strangelets, which means
that $j_0(cos\alpha_{eq})\, = \, constant$ is a reasonable
hypothesis:

\[
P(\alpha_{eq})=4\frac{\alpha_{eq}}{\int_{0}^{\pi/2}\alpha_{eq}\,d\alpha_{eq}}
\]
\\
where the factor $4$ stands for the symmetry in the condition for a
given particle to belong to the loss cone:
$|\alpha_{eq}|<\alpha_{loss\,cone}$ and
$|\pi-\alpha_{eq}|<\alpha_{loss\,cone}$.

Solving the differential equation (\ref{diff}) and obtaining the corresponding
flux inward (in the -$\text{\^e}_r$ direction) for every $L$, it is possible to
determine the mean particle density at a given shell and, in this way, the
trapped strangelet flux.

The results are summarized in Tables 1 and 2 for $L=2$ (ACR belt
location) and $L=1.3$ (location of the maximum of the distribution
function) for the example of strangelets of $A=100$ and energy
corresponding to $R=1 GV$.

\begin{table}
\begin{tabular}{|l|l|l|}
\hline &  $L \, = \, 1.3$ & $L \, = \, 2$ \\
\hline Normal &$1.28 \times 10^{-15}$ & $4.34 \times 10^{-17}$ \\
\hline CFL & $3.95 \times 10^{-14}$ & $1.34 \times 10^{-15}$ \\
\hline
\end{tabular}
\caption{Mean particle flux in units of $part\, cm^{-2}\, s^{-1}\,
sr^{-1}\, (MeV/A)^{-1}$ for strangelet stationary population at $L
\, = \, 1.3$ and  $L \, = \, 2$ calculated with the ``standard"
flux.}
\end{table}

\begin{table}
\begin{tabular}{|l|l|l|}
\hline &  $L \, = \, 1.3$ &  $L \, = \, 2$ \\
\hline Normal &$3.83 \times 10^{-14}$   & $1.3 \times 10^{-15}$ \\
\hline CFL & $1.64 \times 10^{-13}$ & $5.55 \times 10^{-15}$ \\
\hline
\end{tabular}
\caption{Mean particle flux in units of $part\, cm^{-2}\, s^{-1}\,
sr^{-1}\, (MeV/A)^{-1}$ for strangelet stationary population at $L
\, = \, 1.3$ and  $L \, = \, 2$ calculated with the ``improved"
flux.}
\end{table}

The position of the peak of the distribution function in the
geomagnetic field (around $L=1.3$) is quite robust, it does not
appreciably change with the change in the $A/Z$ relation (CFL and
normal strangelets), nor with a change in the energy and baryon
number of the strangelets. This could be a consequence of the
assumption of the power spectral density as being proportional to
$\nu^{-2}$, what renders the diffusion coefficient independent of
the particle energy, therefore modifying the energy of the
particles does not affect their diffusion properties.

We observe that the trapped population is slightly more favored if
strange quark matter is in the CFL state, the difference between
the trapped fluxes for the two species increases with decreasing
energy exponent in the incident flux. It happens due to the
dependence on the baryonic number of the interstellar flux of
strangelets (eq. \ref{flux}). Since the rigidity interval for
stable trapping is the same for both states for it only depends on
the geometrical characteristics of the geomagnetic field, the
difference on the number of particles trapped strongly depends on
the difference in the incoming flux. This dependence of the
integrated flux on the number of baryons that can be expressed as
$F_{ISM}\propto (0.125)^{-1.2}\,A^{-1.667}$ and $F_{ISM}\propto
(0.3)^{-1.2}\,A^{-1.267}$ for normal and CFL strangelets,
respectively. In this way the flux of paired CFL strangelets is
lower than those without pairing, but only for low baryon number
($A < \sim 13$), that is, in a region where it is believed
strangelets are not stable at all. In the stability region the
flux for CFL strangelets is always higher than for normal
strangelets resulting in a higher trapped density. In this way,
the smaller difference seeing between strangelets with and without
pairing for the improved flux when comparing to that for the
standard flux is explained by the smaller difference in the
incoming flux due to the softer dependence on the atomic number
(the standard flux depends on $E^{-2.5}$, which for the same
rigidity depends on the particle's atomic number; instead, the
improved flux depends on $R^{1.8}$ and the analysis was performed
in terms of same rigidity).

Additional considerations are relevant for the fate of a trapped
population of strangelets. It is well-known that the solar wind has
a strong influence on the ACR flux upon the earth. The most abundant
ACR heavy ion, oxygen, shows a strong intensity variation with the
solar cycle, having its interstellar flux of $8-27 \, MeV/nucleon$
lowered up to two orders of magnitude during periods of solar
maximum activity \cite{Biswas}. During solar minimum, the trapped
flux at the earth magnetosphere is of the order of $\sim 5\times
10^{-4} \, particles\, cm^{-2} \, sr^{-1} \, s^{-1} \,
(MeV/nucleon)^{-1}$, corresponding to an enhancement factor of $\sim
15$ \cite{Bobrovskaya}, this experimental value being somewhat below
the theoretical expected one (higher than 25 \cite{Biswas}). The
oxygen component corresponds to about $80 \%$ of the trapped ACR,
while the $C/O$, $N/O$ and $Ne/O$ abundance ratio are $< 0.005$,
$\sim 0.10-0.15$ and $\sim \, 0.02-0.03$, respectively.

With the results obtained in this study, the trapped flux of
strangelets at $L < 2$ would be of order $10^{-14}-10^{-15} \,
particles\, cm^{-2} \, sr^{-1} \, s^{-1} \, (MeV/A)^{-1}$ at
rigidity $R \, = \, 1 \, GV$ for strangelets of baryon number
$A=100$. This represents an enhancement factor for trapped flux in
the regime of steady-state population comparing to the
interstellar flux at the same energy and $A$ of order $10$ and
$10^2$ for strangelets trapped at $L=2$ and $L=1.3$, respectively,
the values for CFL strangelets being of about twice the one for
normal strangelets ($q\sim 5.5$ and $11$, and $q\sim 162$ and
$314$ for CFL and normal strangelets at $L=2$ and $L=1.3$,
respectively). This results show that the strangelet flux could be
as high as a factor 10000 lower than that expected for carbon
during periods of maximum solar activity. Although we did not
consider the solar modulation in our calculations, it would act
significantly over those low energy strangelets \cite{Madsen05},
the region of interest in this study. In this manner, it could
have an important influence, similar to that detected for oxygen,
on the trapped density.

The advantage of a search for trapped strangelets in the geomagnetic
field performed during the solar maximum activity whether they are an
important component of the radiation belt or are to be measured
penetrating the atmosphere towards to surface of the earth
would be the reduced component of ACR, which could reduce dead time losses
in the detectors and possibly render a clearer identification of the primaries.

The proposed and widely accepted model for ACR trapping
\cite{Blake} assumes that the high mass-to-charge ratio of
singly-ionized ACRs enables them to penetrate deeply into the
magnetosphere. ACRs with trajectories near a low altitude mirror
point interact with particles in the upper atmosphere, loosing one
or all their remaining electrons. After stripping, the particle
gyroradius is reduced by a factor of $1/Z$, and the ion can become
stably trapped. As stated above, the results presented here were
obtained assuming fully-ionized strangelets, which have just the
``right'' features to become trapped. However, some fraction of
the strangelets should reach earth's atmosphere with an effective
charge slightly below their atomic number and suffer a process of
interaction similar to ACR's, which is much less dependent on the
pitch angle and other variables. Finally, there is also the
possibility of quasi-stable trapping of ions with energies high
enough not to obey condition (\ref{rho}), but not too high as to
penetrate the magnetosphere without suffering any significant
depletion in their incident direction. These two additional
mechanisms could result in a further {\it increase} in the number
of trapped strangelets.

\section*{Conclusions}

From the analysis presented here we conclude that non-relativistic
strangelets with $A \, < \, \sim \, 10^{3}$ already ionized by
collisions with electrons in the ISM could be stably trapped by
the geomagnetic field. Assuming the existence of a strangelet
contamination in the ISM, its injection in the solar system and
given the geomagnetic geometry and the interaction of the magnetic
field with the solar wind, it looks very likely to have this
radiation belt surrounding the planet. If strangelets are to be a
component of the anomalous cosmic ray belt at $L \, \sim \, 2$, we
have shown that, even considering the approximations taken during
the calculations presented here (which have the main consequence
of averaging the trapped population's behavior), those particles
would be present with an enhancement factor comparing with the
interstellar flux of order $10^{1}$ and if we consider a new
particle belt (a strangelet belt) at $L\,\sim\,1.3$, the
enhancement factor could be as high as order $10^{2}$ in a
stationary population scenario \footnote{It must be noted,
however, that for the ACR belt the theoretical values for the
enhancement factor are somewhat higher than the experimental ones,
a feature that would probably hold for the strangelet belt as
well.}. These exotic baryons could in principle be detectable in
the earth magnetosphere depending on the chosen parameters for
each of the experiments (effective detection area, altitude and
type of orbiting, magnetic field for particle depletion and
others). In addition to the already mentioned capture of almost
fully ionized strangelets, additional trajectories leading to trapping (but not
obeying the adiabatic conditions) may exist, although they must be
calculated numerically, and could enhance even further the trapped
population, though most probably not affecting substantially the
results. Effects that could result in the reduction of the trapped
population are the diffusion driven by electric fields
fluctuations and phenomena directly related to enhanced solar
activity, which though less likely to affect the particles already
trapped at low $L$-shells, could have an influence on the particle
injection in the outer magnetosphere.

Overall, we believe our estimates to be on the conservative side
of the trapped flux, making the search of trapped strangelets a
feasible but difficult task. Needless to say, the detection of
those trapped particles having low $Z/A$ ratio would be extremely
important for determining the properties of cold, dense baryonic
matter.

\begin{acknowledgments}
L.P. acknowledges the financial support received from
the Funda\c c\~ao de Amparo \`a Pesquisa do Estado de
S\~ao Paulo. J.E.H. wishes to acknowledge the CNPq
Agency (Brazil) for partial financial support.
\end{acknowledgments}

\bibliography{geo}

\begin{thebibliography}{59}
\expandafter\ifx\csname natexlab\endcsname\relax\def\natexlab#1{#1}\fi
\expandafter\ifx\csname bibnamefont\endcsname\relax
  \def\bibnamefont#1{#1}\fi
\expandafter\ifx\csname bibfnamefont\endcsname\relax
  \def\bibfnamefont#1{#1}\fi
\expandafter\ifx\csname citenamefont\endcsname\relax
  \def\citenamefont#1{#1}\fi
\expandafter\ifx\csname url\endcsname\relax
  \def\url#1{\texttt{#1}}\fi
\expandafter\ifx\csname urlprefix\endcsname\relax\def\urlprefix{URL }\fi
\providecommand{\bibinfo}[2]{#2}
\providecommand{\eprint}[2][]{\url{#2}}

\bibitem[{\citenamefont{Witten}(1984)}]{Wit}
\bibinfo{author}{\bibfnamefont{E.}~\bibnamefont{Witten}},
  \bibinfo{journal}{Phys.\ Rev. D} \textbf{\bibinfo{volume}{30}},
  \bibinfo{pages}{272} (\bibinfo{year}{1984}).

\bibitem[{\citenamefont{Bodmer}(1971)}]{Bodmer}
\bibinfo{author}{\bibfnamefont{A.}~\bibnamefont{Bodmer}},
  \bibinfo{journal}{Phys.\ Rev. D} \textbf{\bibinfo{volume}{4}},
  \bibinfo{pages}{1601} (\bibinfo{year}{1971}).

\bibitem[{\citenamefont{Chin and Kerman}(1979)}]{Chin}
\bibinfo{author}{\bibfnamefont{S.~A.} \bibnamefont{Chin}} \bibnamefont{and}
  \bibinfo{author}{\bibfnamefont{A.}~\bibnamefont{Kerman}},
  \bibinfo{journal}{Phys.\ Rev. Lett.} \textbf{\bibinfo{volume}{43}},
  \bibinfo{pages}{1292} (\bibinfo{year}{1979}).

\bibitem[{\citenamefont{Terazawa}(1979)}]{Terazawa}
\bibinfo{author}{\bibfnamefont{H.}~\bibnamefont{Terazawa}},
  \bibinfo{journal}{Tokyo U. Report.} pp. \bibinfo{pages}{INS--336}
  (\bibinfo{year}{1979}).

\bibitem[{\citenamefont{Farhi and Jaffe}(1984)}]{Farhi84}
\bibinfo{author}{\bibfnamefont{E.}~\bibnamefont{Farhi}} \bibnamefont{and}
  \bibinfo{author}{\bibfnamefont{L.}~\bibnamefont{Jaffe}},
  \bibinfo{journal}{Phys.\ Rev. D} \textbf{\bibinfo{volume}{30}},
  \bibinfo{pages}{2379} (\bibinfo{year}{1984}).

\bibitem[{\citenamefont{Gilson and Jaffe}(1993)}]{Jaf}
\bibinfo{author}{\bibfnamefont{E.~P.} \bibnamefont{Gilson}} \bibnamefont{and}
  \bibinfo{author}{\bibfnamefont{R.~L.} \bibnamefont{Jaffe}},
  \bibinfo{journal}{Phys.\ Rev. Lett.} \textbf{\bibinfo{volume}{71}},
  \bibinfo{pages}{332} (\bibinfo{year}{1993}).

\bibitem[{\citenamefont{Madsen}({\natexlab{a}})}]{Madsen98}
\bibinfo{author}{\bibfnamefont{J.}~\bibnamefont{Madsen}},
  \eprint{astro-ph/9809032}.

\bibitem[{\citenamefont{Buballa}()}]{NJL}
\bibinfo{author}{\bibfnamefont{M.}~\bibnamefont{Buballa}},
  \eprint{hep-ph/0402234}.

\bibitem[{\citenamefont{Lugones and Horvath}(2002)}]{CFL1}
\bibinfo{author}{\bibfnamefont{G.}~\bibnamefont{Lugones}} \bibnamefont{and}
  \bibinfo{author}{\bibfnamefont{J.}~\bibnamefont{Horvath}},
  \bibinfo{journal}{Phys.\ Rev. D} \textbf{\bibinfo{volume}{66}},
  \bibinfo{pages}{074017} (\bibinfo{year}{2002}).

\bibitem[{\citenamefont{Alford et~al.}(1999)\citenamefont{Alford, Rajagopal,
  and Wilczek}}]{CFL2}
\bibinfo{author}{\bibfnamefont{M.}~\bibnamefont{Alford}},
  \bibinfo{author}{\bibfnamefont{K.}~\bibnamefont{Rajagopal}},
  \bibnamefont{and} \bibinfo{author}{\bibfnamefont{F.}~\bibnamefont{Wilczek}},
  \bibinfo{journal}{Nucl. Phys. B} \textbf{\bibinfo{volume}{537}},
  \bibinfo{pages}{433} (\bibinfo{year}{1999}).

\bibitem[{\citenamefont{Rapp et~al.}(2000)}]{CFL3}
\bibinfo{author}{\bibfnamefont{R.}~\bibnamefont{Rapp}} \bibnamefont{et~al.},
  \bibinfo{journal}{Ann. Phys. (N. Y.)} \textbf{\bibinfo{volume}{280}},
  \bibinfo{pages}{35} (\bibinfo{year}{2000}).

\bibitem[{\citenamefont{Rajagopal and Wilczeck}()}]{CFL4}
\bibinfo{author}{\bibfnamefont{K.}~\bibnamefont{Rajagopal}} \bibnamefont{and}
  \bibinfo{author}{\bibfnamefont{F.}~\bibnamefont{Wilczeck}},
  \bibinfo{note}{for an overview of the CFL state}, \eprint{hep-ph/0011333}.

\bibitem[{\citenamefont{Madsen}(2002)}]{Mads}
\bibinfo{author}{\bibfnamefont{J.}~\bibnamefont{Madsen}},
  \bibinfo{journal}{J.Phys. G} \textbf{\bibinfo{volume}{28}},
  \bibinfo{pages}{1737} (\bibinfo{year}{2002}).

\bibitem[{\citenamefont{Klingenberg}(2001)}]{Klin}
\bibinfo{author}{\bibfnamefont{R.}~\bibnamefont{Klingenberg}},
  \bibinfo{journal}{J. Phys. G} \textbf{\bibinfo{volume}{27}},
  \bibinfo{pages}{475} (\bibinfo{year}{2001}).

\bibitem[{\citenamefont{Finch}()}]{Finch}
\bibinfo{author}{\bibfnamefont{E.}~\bibnamefont{Finch}},
  \eprint{nucl-ex/0605010}.

\bibitem[{\citenamefont{Thomas and Jacobs}()}]{Ions1}
\bibinfo{author}{\bibfnamefont{J.}~\bibnamefont{Thomas}} \bibnamefont{and}
  \bibinfo{author}{\bibfnamefont{P.}~\bibnamefont{Jacobs}},
  \emph{\bibinfo{title}{A guide to the high energy ion experiments}},
  \bibinfo{note}{{UCRL-ID-119181}}.

\bibitem[{\citenamefont{Rusek et~al.}(1996)}]{Ions2}
\bibinfo{author}{\bibfnamefont{A.}~\bibnamefont{Rusek}} \bibnamefont{et~al.}
  (\bibinfo{collaboration}{for the {E886} collaboration}),
  \bibinfo{journal}{Phys. Rev. C} \textbf{\bibinfo{volume}{54}},
  \bibinfo{pages}{R15} (\bibinfo{year}{1996}).

\bibitem[{\citenamefont{{Van Buren}}(1999)}]{Ions3}
\bibinfo{author}{\bibfnamefont{G.}~\bibnamefont{{Van Buren}}}
  (\bibinfo{collaboration}{for the {E864} collaboration}), \bibinfo{journal}{J.
  Phys. G} \textbf{\bibinfo{volume}{25}}, \bibinfo{pages}{411}
  (\bibinfo{year}{1999}).

\bibitem[{\citenamefont{Belz et~al.}(1996)}]{Ions4}
\bibinfo{author}{\bibfnamefont{J.}~\bibnamefont{Belz}} \bibnamefont{et~al.}
  (\bibinfo{collaboration}{for the {BNL E888} collaboration}),
  \bibinfo{journal}{Phys. Rev. Lett.} \textbf{\bibinfo{volume}{76}},
  \bibinfo{pages}{3277} (\bibinfo{year}{1996}).

\bibitem[{\citenamefont{Dittus et~al.}(1995)}]{Ions5}
\bibinfo{author}{\bibfnamefont{F.}~\bibnamefont{Dittus}} \bibnamefont{et~al.}
  (\bibinfo{collaboration}{for the {NA52} collaboration}), in
  \emph{\bibinfo{booktitle}{International Conference on Strangeness in Hadronic
  Matter}}, edited by
  \bibinfo{editor}{\bibfnamefont{J.}~\bibnamefont{Rafelski}}
  (\bibinfo{publisher}{American Institute of Physics, New York (AIP 340)},
  \bibinfo{year}{1995}), p.~\bibinfo{pages}{24}.

\bibitem[{\citenamefont{Appelquist et~al.}(1996)}]{Ions6}
\bibinfo{author}{\bibfnamefont{G.}~\bibnamefont{Appelquist}}
  \bibnamefont{et~al.}, \bibinfo{journal}{Phys. Rev. Lett.}
  \textbf{\bibinfo{volume}{76}}, \bibinfo{pages}{3907} (\bibinfo{year}{1996}).

\bibitem[{\citenamefont{Ambrosini et~al.}(1996)}]{Ions7}
\bibinfo{author}{\bibfnamefont{G.}~\bibnamefont{Ambrosini}}
  \bibnamefont{et~al.}, \bibinfo{journal}{Nucl. Phys.}
  \textbf{\bibinfo{volume}{A610}}, \bibinfo{pages}{306c}
  (\bibinfo{year}{1996}).

\bibitem[{\citenamefont{Klingenberg}(1999)}]{Ions8}
\bibinfo{author}{\bibfnamefont{R.}~\bibnamefont{Klingenberg}},
  \bibinfo{journal}{J. Phys. G} \textbf{\bibinfo{volume}{25}},
  \bibinfo{pages}{R273} (\bibinfo{year}{1999}).

\bibitem[{\citenamefont{Arsenescu et~al.}(2002)}]{Arsen}
\bibinfo{author}{\bibfnamefont{R.}~\bibnamefont{Arsenescu}}
  \bibnamefont{et~al.}, \bibinfo{journal}{New J. Phys.}
  \textbf{\bibinfo{volume}{4}}, \bibinfo{pages}{96} (\bibinfo{year}{2002}).

\bibitem[{\citenamefont{{De Rujula} and Glashow}(1984)}]{derujula84}
\bibinfo{author}{\bibfnamefont{A.}~\bibnamefont{{De Rujula}}} \bibnamefont{and}
  \bibinfo{author}{\bibfnamefont{S.~L.} \bibnamefont{Glashow}},
  \bibinfo{journal}{Nature} \textbf{\bibinfo{volume}{312}},
  \bibinfo{pages}{734} (\bibinfo{year}{1984}).

\bibitem[{\citenamefont{Bruegger et~al.}(1989)}]{bruegger89}
\bibinfo{author}{\bibfnamefont{M.}~\bibnamefont{Bruegger}}
  \bibnamefont{et~al.}, \bibinfo{journal}{Nature}
  \textbf{\bibinfo{volume}{337}}, \bibinfo{pages}{434} (\bibinfo{year}{1989}).

\bibitem[{\citenamefont{Isaac et~al.}()}]{isaac98}
\bibinfo{author}{\bibfnamefont{M.~C.~P.} \bibnamefont{Isaac}}
  \bibnamefont{et~al.}, \eprint{astro-ph/9806147}.

\bibitem[{\citenamefont{Lu et~al.}()}]{Lu04}
\bibinfo{author}{\bibfnamefont{Z.-T.} \bibnamefont{Lu}} \bibnamefont{et~al.},
  \eprint{nucl-ex/0402015}.

\bibitem[{\citenamefont{Bjorken and McLerran}(1979)}]{Data1}
\bibinfo{author}{\bibfnamefont{J.~D.} \bibnamefont{Bjorken}} \bibnamefont{and}
  \bibinfo{author}{\bibfnamefont{L.}~\bibnamefont{McLerran}},
  \bibinfo{journal}{Phys.\ Rev. D} \textbf{\bibinfo{volume}{20}},
  \bibinfo{pages}{2353} (\bibinfo{year}{1979}).

\bibitem[{\citenamefont{Rybczynski
  et~al.}({\natexlab{a}})\citenamefont{Rybczynski, Wlodarczyk, and
  Wilk}}]{Data2}
\bibinfo{author}{\bibfnamefont{M.}~\bibnamefont{Rybczynski}},
  \bibinfo{author}{\bibfnamefont{Z.}~\bibnamefont{Wlodarczyk}},
  \bibnamefont{and} \bibinfo{author}{\bibfnamefont{G.}~\bibnamefont{Wilk}},
  \eprint{hep-ph/0109225}.

\bibitem[{\citenamefont{Saito et~al.}(1990)}]{Data3}
\bibinfo{author}{\bibfnamefont{T.}~\bibnamefont{Saito}} \bibnamefont{et~al.},
  \bibinfo{journal}{Phys.\ Rev. Lett.} \textbf{\bibinfo{volume}{65}},
  \bibinfo{pages}{2094} (\bibinfo{year}{1990}).

\bibitem[{\citenamefont{Rybczynski
  et~al.}({\natexlab{b}})\citenamefont{Rybczynski, Wlodarczyk, and Wilk}}]{Pol}
\bibinfo{author}{\bibfnamefont{M.}~\bibnamefont{Rybczynski}},
  \bibinfo{author}{\bibfnamefont{Z.}~\bibnamefont{Wlodarczyk}},
  \bibnamefont{and} \bibinfo{author}{\bibfnamefont{G.}~\bibnamefont{Wilk}},
  \eprint{hep-ph/0410065}.

\bibitem[{\citenamefont{Choutko}(2003)}]{Boiko}
\bibinfo{author}{\bibfnamefont{V.}~\bibnamefont{Choutko}}
  (\bibinfo{collaboration}{for the AMS-01 Collaboration}), in
  \emph{\bibinfo{booktitle}{Proc. 28th Internat. Cosmic Ray Conf., Tsukuba,
  Japan}} (\bibinfo{publisher}{Universal Academic Press, Tokyo},
  \bibinfo{year}{2003}), p. \bibinfo{pages}{1765}.

\bibitem[{\citenamefont{Horvath and Paulucci}(2006)}]{PaulucciHorvath}
\bibinfo{author}{\bibfnamefont{J.~E.} \bibnamefont{Horvath}} \bibnamefont{and}
  \bibinfo{author}{\bibfnamefont{L.}~\bibnamefont{Paulucci}},
  \bibinfo{journal}{J. Phys. G} \textbf{\bibinfo{volume}{32}},
  \bibinfo{pages}{B13} (\bibinfo{year}{2006}).

\bibitem[{\citenamefont{Kalogera et~al.}(2004)}]{DP}
\bibinfo{author}{\bibfnamefont{V.}~\bibnamefont{Kalogera}}
  \bibnamefont{et~al.}, \bibinfo{journal}{Astrophys. J. Lett.}
  \textbf{\bibinfo{volume}{601}}, \bibinfo{pages}{L179} (\bibinfo{year}{2004}).

\bibitem[{\citenamefont{Kluzniak and Lee}(2002)}]{Wod}
\bibinfo{author}{\bibfnamefont{W.}~\bibnamefont{Kluzniak}} \bibnamefont{and}
  \bibinfo{author}{\bibfnamefont{W.}~\bibnamefont{Lee}},
  \bibinfo{journal}{MNRAS} \textbf{\bibinfo{volume}{335}}, \bibinfo{pages}{L29}
  (\bibinfo{year}{2002}).

\bibitem[{\citenamefont{Limousin et~al.}()\citenamefont{Limousin,
  Gondek-Rosinska, and Gourgoulhon}}]{Coalescencia}
\bibinfo{author}{\bibfnamefont{F.}~\bibnamefont{Limousin}},
  \bibinfo{author}{\bibfnamefont{D.}~\bibnamefont{Gondek-Rosinska}},
  \bibnamefont{and}
  \bibinfo{author}{\bibfnamefont{E.}~\bibnamefont{Gourgoulhon}},
  \eprint{gr-qc/0411127}.

\bibitem[{\citenamefont{Benvenuto and Horvath}(1989)}]{Mac}
\bibinfo{author}{\bibfnamefont{O.~G.} \bibnamefont{Benvenuto}}
  \bibnamefont{and} \bibinfo{author}{\bibfnamefont{J.~E.}
  \bibnamefont{Horvath}}, \bibinfo{journal}{Phys.\ Rev. Lett.}
  \textbf{\bibinfo{volume}{63}}, \bibinfo{pages}{716} (\bibinfo{year}{1989}).

\bibitem[{\citenamefont{Medina-Tanco and Horvath}(1996)}]{Gustavo}
\bibinfo{author}{\bibfnamefont{G.~A.} \bibnamefont{Medina-Tanco}}
  \bibnamefont{and} \bibinfo{author}{\bibfnamefont{J.~E.}
  \bibnamefont{Horvath}}, \bibinfo{journal}{Astrophys. J.}
  \textbf{\bibinfo{volume}{464}}, \bibinfo{pages}{364} (\bibinfo{year}{1996}).

\bibitem[{\citenamefont{Madsen}(2005)}]{Madsen05}
\bibinfo{author}{\bibfnamefont{J.}~\bibnamefont{Madsen}},
  \bibinfo{journal}{Phys.\ Rev. D} \textbf{\bibinfo{volume}{71}},
  \bibinfo{pages}{014026} (\bibinfo{year}{2005}).

\bibitem[{\citenamefont{Rosu}()}]{Rosu}
\bibinfo{author}{\bibfnamefont{H.~C.} \bibnamefont{Rosu}},
  \eprint{hep-ph/9410028}.

\bibitem[{AMS()}]{AMS}
\bibinfo{note}{Home page of AMS experiment},
  \urlprefix\url{http://ams.cern.ch}.

\bibitem[{\citenamefont{Madsen}({\natexlab{b}})}]{AMS2}
\bibinfo{author}{\bibfnamefont{J.}~\bibnamefont{Madsen}},
  \eprint{hep-ph/0111417}.

\bibitem[{\citenamefont{Rajagopal and Wilczek}(2001)}]{Rajagopal}
\bibinfo{author}{\bibfnamefont{K.}~\bibnamefont{Rajagopal}} \bibnamefont{and}
  \bibinfo{author}{\bibfnamefont{F.}~\bibnamefont{Wilczek}},
  \bibinfo{journal}{Phys. Rev. Lett.} \textbf{\bibinfo{volume}{86}},
  \bibinfo{pages}{3492} (\bibinfo{year}{2001}).

\bibitem[{\citenamefont{Madsen and Larsen}(2003)}]{Madsen03}
\bibinfo{author}{\bibfnamefont{J.}~\bibnamefont{Madsen}} \bibnamefont{and}
  \bibinfo{author}{\bibfnamefont{J.~M.} \bibnamefont{Larsen}},
  \bibinfo{journal}{Phys.\ Rev. Lett.} \textbf{\bibinfo{volume}{90}},
  \bibinfo{pages}{121102} (\bibinfo{year}{2003}).

\bibitem[{\citenamefont{Lattes and Fujimoto}(1980)}]{Lattes}
\bibinfo{author}{\bibfnamefont{C.}~\bibnamefont{Lattes}} \bibnamefont{and}
  \bibinfo{author}{\bibfnamefont{K.}~\bibnamefont{Fujimoto}},
  \bibinfo{journal}{Phys. Repts.} \textbf{\bibinfo{volume}{65}},
  \bibinfo{pages}{151} (\bibinfo{year}{1980}).

\bibitem[{\citenamefont{Werner and Salpeter}(1969)}]{Werner}
\bibinfo{author}{\bibfnamefont{M.~W.} \bibnamefont{Werner}} \bibnamefont{and}
  \bibinfo{author}{\bibfnamefont{E.~E.} \bibnamefont{Salpeter}},
  \bibinfo{journal}{MNRAS} \textbf{\bibinfo{volume}{145}}, \bibinfo{pages}{249}
  (\bibinfo{year}{1969}).

\bibitem[{\citenamefont{Massey and Burhop}(1952)}]{Massey}
\bibinfo{author}{\bibfnamefont{H.~S.} \bibnamefont{Massey}} \bibnamefont{and}
  \bibinfo{author}{\bibfnamefont{E.~H.~S.} \bibnamefont{Burhop}},
  \emph{\bibinfo{title}{Electronic and Ionic Impact Phenomena}}
  (\bibinfo{publisher}{Oxford University Press}, \bibinfo{year}{1952}).

\bibitem[{\citenamefont{Bethe and Salpeter}(1957)}]{Bethe}
\bibinfo{author}{\bibfnamefont{H.~A.} \bibnamefont{Bethe}} \bibnamefont{and}
  \bibinfo{author}{\bibfnamefont{E.~E.} \bibnamefont{Salpeter}},
  \emph{\bibinfo{title}{Handbuch der Physik}}, vol.~\bibinfo{volume}{35}
  (\bibinfo{publisher}{Springer--Verlag Berlin}, \bibinfo{year}{1957}).

\bibitem[{\citenamefont{Bohr}(1948)}]{Bohr48}
\bibinfo{author}{\bibfnamefont{N.}~\bibnamefont{Bohr}}, \bibinfo{journal}{Kgl.
  Dansk. Vid. Selsk.} \textbf{\bibinfo{volume}{18}}, \bibinfo{pages}{8}
  (\bibinfo{year}{1948}).

\bibitem[{\citenamefont{Stoermer}(1955)}]{Stormer}
\bibinfo{author}{\bibfnamefont{C.}~\bibnamefont{Stoermer}},
  \emph{\bibinfo{title}{The polar aurora}} (\bibinfo{publisher}{Claredon Press,
  Oxford}, \bibinfo{year}{1955}).

\bibitem[{\citenamefont{Adams et~al.}(1981)\citenamefont{Adams, Silberberg, and
  Tsao}}]{Cut}
\bibinfo{author}{\bibfnamefont{J.~H.} \bibnamefont{Adams}},
  \bibinfo{author}{\bibfnamefont{R.}~\bibnamefont{Silberberg}},
  \bibnamefont{and} \bibinfo{author}{\bibfnamefont{C.~H.} \bibnamefont{Tsao}},
  \bibinfo{type}{Tech. Rep.} \bibinfo{number}{NRL Memorandum Report 4506},
  \bibinfo{institution}{Naval Research Laboratory, Washington DC 20375-500,
  USA} (\bibinfo{year}{1981}), \bibinfo{note}{{S}ee also, for example, D. F.
  Smart and M. A. Shea, Adv. Space Res. {\bf 14}(10), 787 (1994)}.

\bibitem[{\citenamefont{Tylka}(1993)}]{Tylka}
\bibinfo{author}{\bibfnamefont{A.~J.} \bibnamefont{Tylka}}, in
  \emph{\bibinfo{booktitle}{Proc. 23rd Internat. Cosmic Ray Conf., Calgary,
  Canada}} (\bibinfo{year}{1993}), vol.~\bibinfo{volume}{3}, p.
  \bibinfo{pages}{436}.

\bibitem[{\citenamefont{Walt}(1994)}]{Walt}
\bibinfo{author}{\bibfnamefont{M.}~\bibnamefont{Walt}},
  \emph{\bibinfo{title}{Introduction to geomagnetically trapped radiation}}
  (\bibinfo{publisher}{Cambridge Press}, \bibinfo{year}{1994}).

\bibitem[{\citenamefont{Baumjohann}()}]{Baumjohann}
\bibinfo{author}{\bibfnamefont{W.}~\bibnamefont{Baumjohann}},
  \bibinfo{note}{private communication}.

\bibitem[{\citenamefont{Biswas}(1996)}]{Biswas}
\bibinfo{author}{\bibfnamefont{S.}~\bibnamefont{Biswas}},
  \bibinfo{journal}{Space Science Rev.} \textbf{\bibinfo{volume}{75}},
  \bibinfo{pages}{423} (\bibinfo{year}{1996}).

\bibitem[{\citenamefont{Bobrovskaya et~al.}(2001)}]{Bobrovskaya}
\bibinfo{author}{\bibfnamefont{V.~V.} \bibnamefont{Bobrovskaya}}
  \bibnamefont{et~al.}, \bibinfo{journal}{Cosmic Research}
  \textbf{\bibinfo{volume}{39}}, \bibinfo{pages}{98} (\bibinfo{year}{2001}).

\bibitem[{\citenamefont{Blake and Friesen}(1977)}]{Blake}
\bibinfo{author}{\bibfnamefont{J.~B.} \bibnamefont{Blake}} \bibnamefont{and}
  \bibinfo{author}{\bibfnamefont{L.~M.} \bibnamefont{Friesen}}, in
  \emph{\bibinfo{booktitle}{Proc. 15th Internat. Cosmic Ray Conf., Plovdiv,
  Bulgaria}} (\bibinfo{year}{1977}), vol.~\bibinfo{volume}{2}, p.
  \bibinfo{pages}{341}.

\bibitem[{\citenamefont{Falthammar}(1968)}]{electric}
\bibinfo{author}{\bibfnamefont{C.-G.} \bibnamefont{Falthammar}}, in
  \emph{\bibinfo{booktitle}{Earth's Particles and Fields}}
  (\bibinfo{publisher}{Ed.B.M. McCormac, Reinhold Book Co.},
  \bibinfo{year}{1968}), pp. \bibinfo{pages}{157--169}.

\end{thebibliography}

\end{document}